# A standardized file format and open-source analysis framework for Brillouin microscopy data

Carlo Bevilacqua[1,*], Sebastian Hambura[1,*], Pierre Bouvet[2,*], Salvatore La Cavera III[3], Kareem Elsayad[2,*,#], and Robert Prevedel[1,4,*,#]

[1]Cell Biology and Biophysics Unit, European Molecular Biology Laboratory, Heidelberg, Germany.
[2]Center for Anatomy and Cell Biology, Medical University of Vienna, Vienna, Austria.
[3]Optics & Photonics Group, Faculty of Engineering, University of Nottingham, Nottingham, UK.
[4]German Center for Lung Research (DZL), Heidelberg, Germany.

*Contributed equally. #Corresponding authors.

**Brillouin microscopy is rapidly emerging as a powerful technique for imaging the mechanical properties of biological specimens in a label-free, non-contact manner. We present a standardized file format and open-source tools to facilitate the uptake and analysis of Brillouin microscopy related data and to unify this growing field.**

Brillouin microscopy allows for 3D mapping of the viscoelasticity in living cells and tissue with a high spatio-temporal resolution[1,2]. It is based on inelastic scattering – so-called Brillouin Light Scattering (BLS)[3] – that results from the interaction of photons with propagating thermally induced or stimulated MHz-GHz acoustic waves (*acoustic phonons*). The frequency or time shift of this scattered light relative to the elastic scattered light, and its spectral-width or attenuation, are directly related to the material's longitudinal elastic and viscous moduli[1,3,4]. For certain implementations, the intensity and angle dependence of the BLS scattering can also be used to estimate the mass density[5] and the refractive index[6,7]. Over the past decade, it has found increasing applications for studying cell and tissue mechanics in development and disease, as well as for the physical characterization of biomaterials, largely on accounts of improvements in the required ultra-high resolution spectrometer designs[1,2,8–10].

Despite its growing adoption, the Brillouin microscopy community faces a critical bottleneck: the lack of standardized methods for storing, sharing, and analyzing spectral data and associated metadata. This limits reproducibility and hampers cross-study and cross-lab comparisons — challenges that become increasingly acute as the field as well as number and diversity of users increases[11].

In Brillouin images, each spatial position (voxel) is associated with a spectrum over a finite spectral range (typically a few to 10s of GHz) in the vicinity of the probing laser frequency (**Fig. 1a**). What is generally of interest is the position, shape, intensity and width of the spectrum which can be composed of a single Brillouin peak, or in case of heterogenous materials, combination of multiple peaks, which are fitted using established models to obtain values for the viscoelastic parameters. The useful information is generally contained in the rendered spatial maps of these fitted parameters for each voxel. The diverse modalities for measuring and analyzing the Brillouin spectrum can lead to significant variability in reported values for the same specimen[11]. It is thus a high priority to establish standardized analysis protocols, pipelines and file formats to increase reproducibility and enable meaningful comparisons, especially since relevant parameter changes are often only a few percent.

To address this need, we propose a standardized file format and accompanying open-source software suite for Brillouin microscopy. Our framework supports data from the diverse range of Brillouin microscope implementations — including spontaneous and stimulated scattering, confocal and line-scanning modalities, and time- or frequency-domain detection — offering a unifying and intuitive, easy-to-use solution for both custom-built as well as commercial systems. To showcase this ability, we provide example datasets from diverse Brillouin modalities (**Supplementary Information**).

We advocate for the use of a standardized file format, **.brim**, a hierarchical, cloud-compatible format based on Zarr v3 that is already gaining traction in bioimaging and genomics[12]. Zarr provides efficient storage of large multi-dimensional arrays alongside rich metadata and supports both local and remote data access. Our proposed structure accommodates spectral data, processed outputs (e.g. Brillouin shift and linewidth maps), and relevant acquisition and optional post-processing metadata, together with the corresponding raw data. The design ensures machine readability and future-proofing through explicit versioning and extensible subtypes. We further highlight a complementary file format, **.brimX,** for storing complex multi-parameter biophysical studies that generalize beyond individual or a single series of spatial maps/images. The .brimX format can save entire experiments or projects (e.g. measurements series of different mutants, conditions, etc.) in a single self-contained annotated file, from which select .brim files can easily be extracted using a provided open-source conversion library (*BrimConverter*). This is particularly useful for archiving purposes and sharing large multi-parameter studies in collaborative projects. The adoption of these two common formats can thus broadly cater to all of the Brillouin light-scattering spectroscopy community.

To promote adoption and ease of use, we developed two complementary Python-based packages with extensive documentation (see **Supplementary Information** and **Code availability**):

1. ***Brimfile***, optimized for bioimaging use cases, supports straightforward saving of spectral data as well as analysis/processing results into the standardized .brim format together with the relevant metadata; conversely it also facilitates reading such data from an existing .brim file.

2. ***HDF5_BLS***, tailored to seamlessly integrate into existing Python workflows to create and export standardized HDF5 file formats for all BLS modalities.

These two packages, optimised for both efficiency and flexibility respectively, complement each other through a conversion library integrated into both called *BrimConverter*, and permits reformatting and restructuring between .brim and .brimX files. To allow the use of established image analysis tools on the derived spatial maps (eg. shift, width, amplitude, etc.) both *brimfile* and *HDF5_BLS* support export to widely adopted formats such as OME-TIFF[13]. To further facilitate uptake, we also developed a [Napari plugin](#) to open such maps directly from a .brim file (see **Supplementary Information**).

For spectra visualization and analysis we developed ***BrimView***, a modular, browser-based GUI (developed using Panel/Holoviz) that facilitates the visual representation of derived spatial maps of e.g. shift, width, amplitude, etc., together with the corresponding spectral data. BrimView integrates a package we developed called ***Treat*** which enables processing of the spectra, supporting multiple fitting models (e.g., Lorentzian, Damped Harmonic Oscillator, Voigt) of single or overlapping peaks, and including the option to account for the instrument spectral response function. All the metadata required to make the processing reproducible can be stored in the file. Furthermore, *BrimView* can be run online without requiring local software installation, lowering the adoption barrier for non-expert users. Additionally, the possibility of loading files directly from the cloud further fosters data sharing in the community.

All combined, our infrastructure enables reproducible, transparent analysis workflows and fosters FAIR (Findable, Accessible, Interoperable, Reusable) data practices[14]. Crucially, the standardized format ensures that essential metadata—such as optical configuration, calibration parameters, and fitting assumptions—are preserved and shareable, addressing a major shortcoming in the field, where such information in current publications is often absent or non-machine-readable[11]. We envision this effort as a starting point for a community-driven Brillouin data management ecosystem. Our software is fully open-source, and the file format specifications are hosted on GitHub to encourage feedback, extensions, and broad adoption (see **Supplementary Information**). We believe this standardization initiative is a critical step for Brillouin microscopy to transition from niche technology to a mainstream bioimaging tool. It will facilitate data sharing and comparative analysis across labs, accelerate method development, and increase the impact and accessibility of published studies. We welcome collaboration from the broader microscopy and bioimaging communities as we work toward integration with existing standards, including the OME data model[15] and broader microscopy metadata efforts, such as REMBI[16]. We note that the presented framework can, with suitable customizations, also be adopted for other multi-dimensional hyperspectral datasets beyond Brillouin microscopy as well as for correlating Brillouin microscopy studies with complementary techniques (Raman, fluorescence, etc.). Finally, with the first commercial Brillouin microscopes suitable for routine life-science applications coming to the market, we emphasise the timeliness of establishing a broad adoption of open-source standards, and strongly encourage compatibility with the proposed platform by commercial stakeholders.

**Correspondence** should be addressed to Kareem.Elsayad@meduniwien.ac.at and prevedel@embl.de.

**Data availability.** Sample data are available at https://storage.googleapis.com/brim-example-files. Exemplary imaging data presented in figures are available upon request.

**Code availability.**

| Software | Source code | Documentation |
|---|---|---|
| *BrimView* (https://biobrillouin.org/bri | https://github.com/prevedel-lab/BrimView | |
| mview/) | | |
| brimfile | https://github.com/prevedel-lab/brimfile | https://prevedel-lab.github.io/brimfile/ |
| HDF5_BLS | https://github.com/bio-brillouin/HDF5_BLS | https://github.com/bio-brillouin/HDF5_BLS/blob/main/guides/Tutorial/Tutorial.pdf |
| HDF5_BLS_treat | https://github.com/bio-brillouin/HDF5_BLS/tree/main/HDF5_BLS_treat | https://hdf5-bls.readthedocs.io/en/latest/source/modules.html |

Queries and feedback can be provided by creating an issue on GitHub. Documentation is provided in the **Supplementary Information** (and is kept up to date at https://prevedel-lab.github.io/brimfile/brimfile.html).

## Acknowledgements

The authors wish to thank Jean-Karim Hériché (EMBL) for helpful discussions. C.B., S.B. and R.P. acknowledge funding from the EMBL, an ERC Consolidator Grant (grant no. 864027, Brillouin4Life) and the German Center for Lung Research (DZL). P.B. and K.E. acknowledge funding from the Medical University of Vienna, Austrian Science Foundation FWF (grant no. P34783) and funding from the EU - Recovery and Resilience Facility (RRF). S.L. acknowledges funding from the Royal Academy of Engineering Research Fellowships scheme (grant no. RF-2324-23-223) and the Nottingham Research Fellowship scheme.

## Conflicts of Interest

R.P. is advisor of CellSense Technologies, which makes Brillouin microscopes commercially available. K.E. serves as a scientific advisor for Specto Srl, which manufactures BLS spectrometers.

**Figure 1**

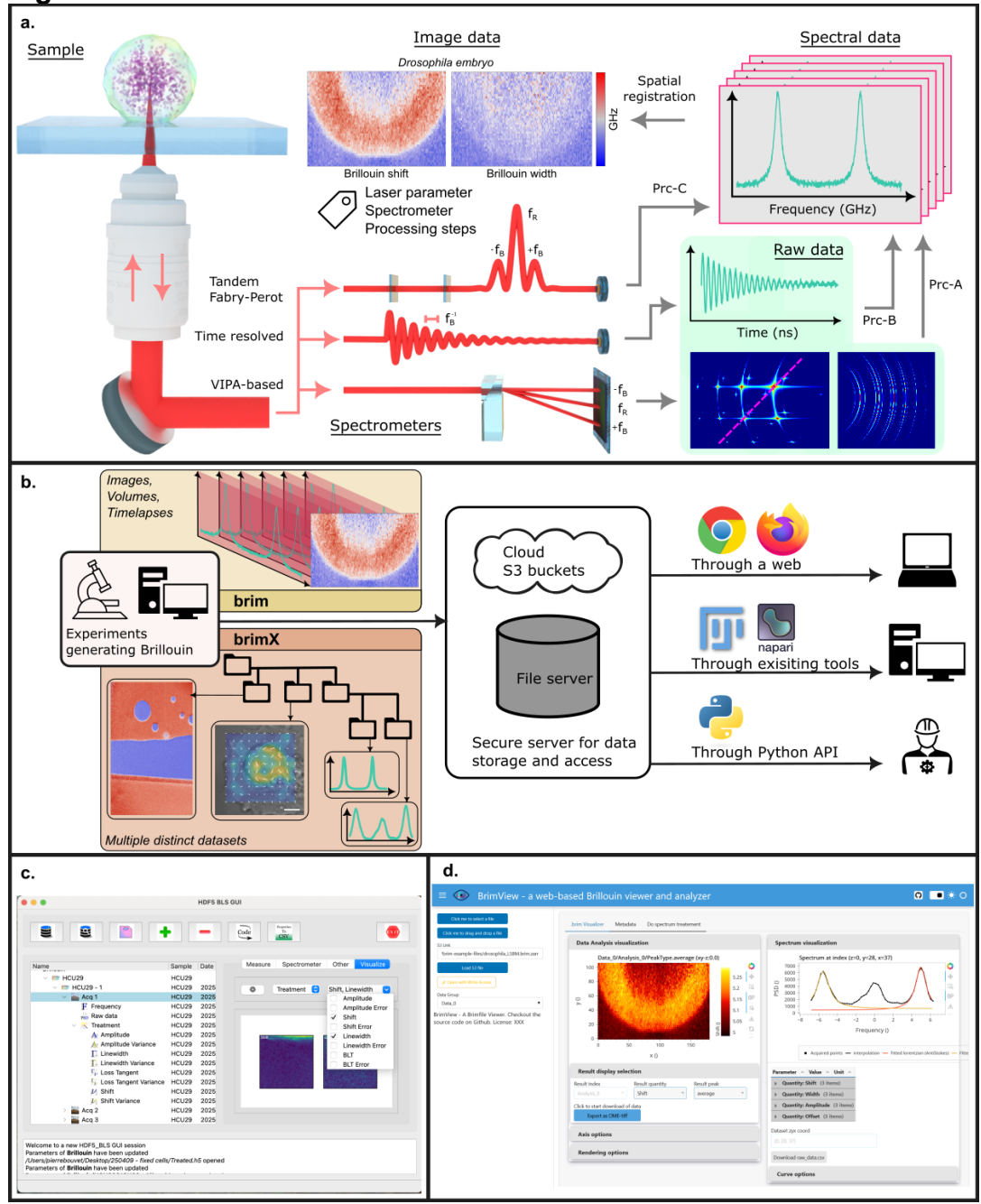

**Fig. 1: A standardized file format and open-source analysis framework for Brillouin microscopy data**. **a)** Overview of Brillouin light scattering data acquisition: A sample is illuminated by a narrowband laser, and the Brillouin scattered light is detected through a high-resolution spectrometer. Different spectrometer modalities produce different raw data that has to be converted into a calibrated and normalized spectrum through dedicated processing pipelines (Prc). This can then be processed into images and mappings of physical properties. **b)** Overview of the Brillouin data analysis pipeline: Brillouin data is saved either as individual .brim images or combined with others into a .brimX (for multi variable experiments). This file is then stored on a data server, from which users can access the recorded data either through cloud-native solution, e.g. S3 buckets, or more traditional means, e.g. a standard file server. The data can be opened through a custom, installationless web-app, through different existing software or with a python library/API. **c)** The HDF5-BLS GUI allows user-friendly creation of .brimX files. **d)** Overview of the *BrimView* GUI, a web-based app that opens .brim files and enables real-time data processing, exploration and visualization.

# Supplementary Information

**A standardized file format and open-source analysis framework for Brillouin microscopy data**

Carlo Bevilacqua[1,*], Sebastian Hambura[1,*], Pierre Bouvet[2,*], Salvatore La Cavera III[3], Kareem Elsayad[2,*,#], and Robert Prevedel[1,4,*,#]

[1]Cell Biology and Biophysics Unit, European Molecular Biology Laboratory, Heidelberg, Germany.
[2]Center for Anatomy and Cell Biology, Medical University of Vienna, Vienna, Austria.
[3]Optics & Photonics Group, Faculty of Engineering, University of Nottingham, Nottingham, UK.
[4]German Center for Lung Research (DZL), Heidelberg, Germany.

*Contributed equally. #Corresponding authors.

## Brim file format for imaging data

We defined a new file format - .brim (= **Br**illouin **im**aging) - with the intent of associating spatial maps to their corresponding spectral information and metadata in a well-defined yet general and flexible fashion. After initially considering HDF5 as a container, we ultimately decided to advocate for the use of the Zarr file format, which shares the same features as HDF5 (mainly allowing the storage of N-dimensional arrays in a hierarchical structure, together with relevant metadata), while being optimized for cloud storage and parallel I/O.

The complete structure of the proposed file format can be found at https://github.com/prevedel-lab/Brillouin-standard-file/blob/main/docs/brim_file_specs.md.

Briefly, a single file could contain multiple images, each of them stored under a 'Data_n' group. The idea is to support multiple timepoints in a timelapse or images of the same sample measured under different conditions, e.g. temperature, osmolarity, drug concentration, etc.. (refer to the 'Conditions' attribute). Each 'Data_n' must contain a 'PSD' (Power Spectral Density) and 'Frequency' array storing the spectral data and the corresponding frequency. The 'PSD' array can optionally store "multidimensional" spectra, which can be useful when each individual spectrum depends on one (or more) external parameters (e.g. angle resolved measurements). Additionally, each 'Data_n' group must contain a 'Scanning' group which associates each spectrum to a spatial position in a flexible way (see the GitHub repo for further information). Optionally one (or more) 'Analysis_m' group(s) can be present in a 'Data_n' group. They contain the relevant quantities which were extracted from the spectra (e.g. shift, width, etc..) under different analysis pipelines (e.g. different lineshape or fitting procedures).

The metadata, defined at https://github.com/prevedel-lab/Brillouin-standard-file/blob/main/docs/brim_file_metadata.md, are stored in the 'Metadata' group and can optionally be redefined in each 'Data_n' group, if different.

By design additional groups, arrays or attributes can be added to a brim file without breaking compatibility with the specs, thus allowing to store any additional data/metadata which might be relevant. The major drawback of this approach is that the additional data is not recognizable by standard softwares. To mitigate this issue we introduced the concept of 'subtypes': this allows defining additional "features" in a .brim file which are not general enough to be included in the definition of a standard brim file but might be common to multiple instruments sharing similar technical implementation (e.g. VIPA-based spectrometer, stimulated Brillouin scattering, etc…). New subtypes should be defined in https://github.com/prevedel-lab/Brillouin-standard-file/blob/main/docs/brim_file_subtypes.md, so that they are available to the community and could be included in standard softwares.

Input from the community is encouraged by creating a GitHub issue https://github.com/prevedel-lab/Brillouin-standard-file/issues

## Brimfile Python library for working with brim files

To easily save data to the brim file format or read from it, we developed a Python library called *brimfile* (https://pypi.org/project/brimfile/), licensed under LGPL-3.0. The full documentation of the library can be found at https://prevedel-lab.github.io/brimfile/brimfile.html
Briefly the library can be installed from PyPI by running the following:
```
pip install brimfile
```

The *brimfile* library is designed in an object-oriented fashion and exposes objects which reflect the structure of the brim file. The user can access the data, metadata and analysis results through their corresponding classes.

- File: represents a brim file, which can be opened or created.
- Data: represents a data group in the brim file, which contains the spectral data and metadata.
- Metadata: represents the metadata associated to a data group (or to the whole file).
- AnalysysResults: represents the results of the analysis of the spectral data.

As an example the following code saves the spectral data together with the corresponding shift and width (generated by a function `generate_data()` not reported here for conciseness) and some metadata to a new brim file::
```python
from brimfile import File, Data, Metadata
from datetime import datetime

filename = 'path/to/your/file.brim.zip'

f = File.create(filename)

PSD, freq_GHz, (dz,dy,dx), shift_GHz, width_GHz = generate_data()

d0 = f.create_data_group(PSD, freq_GHz, (dz,dy,dx), name='test1')

# Create the metadata
Attr = Metadata.Item
datetime_now = datetime.now().isoformat()
temp = Attr(22.0, 'C')
md = d0.get_metadata()

md.add(Metadata.Type.Experiment, {'Datetime':datetime_now, 'Temperature':temp})
md.add(Metadata.Type.Optics, {'Wavelength':Attr(660, 'nm')})
# Add some metadata to the local data group
```

```
    temp = Attr(37.0, 'C')
    md.add(Metadata.Type.Experiment, {'Temperature':temp}, local=True)

    # create the analysis results
    ar = d0.create_analysis_results_group({'shift':shift_GHz, 'shift_units': 'GHz',
                            'width': width_GHz, 'width_units': 'Hz'},
                           {'shift':shift_GHz, 'shift_units': 'GHz',
                            'width': width_GHz, 'width_units': 'Hz'},
                           name = 'test1_analysis')
    f.close()
```

## Napari plugin to open a brimfile

As established image analysis/exploration tools don't readily support our hyperspectral data format, we developed a Napari plugin, called *brillouin-imaging*, which can be installed from the Napari hub (see **SI Fig. 1**). The plugin includes sample data (stored in a S3 bucket) which can be opened to trial it.

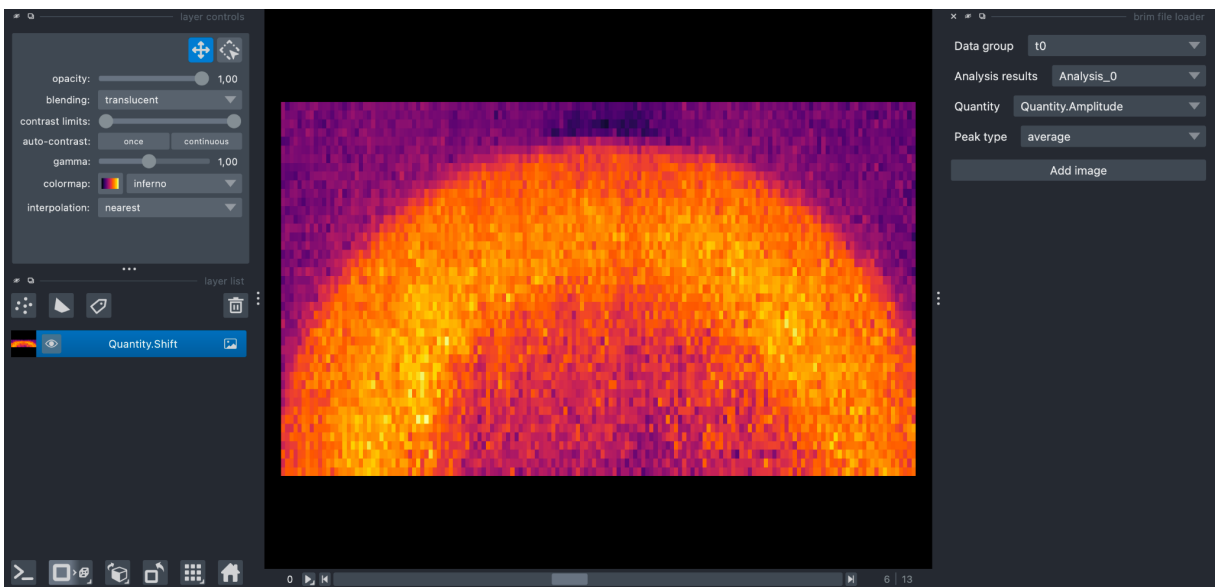

**SI Fig. 1: Overview of the brillouin-imaging Napari plugin**. When a .brim file is opened, a widget is shown on the right of the Napari GUI. Through several dropdown menus the user can select which data group, derived spatial map (e.g. shift, width, etc..), peak type (Stokes or anti-Stokes) to load as new image layer in the Napari viewer.

## Visualization and analysis software

To aid in the analysis and visualization of Brillouin microscopy data, we developed ***BrimView***: a browser-based application that enables interactive visualization and analysis of multidimensional Brillouin microscopy datasets stored in the .brim file format. It is designed to support common workflows used by biologists and other experimental users without requiring software installation or programming knowledge.

*BrimView* runs entirely in modern web browsers and supports both local data access and remote loading from S3-compatible object storage. This enables flexible deployment in both online and offline environments.

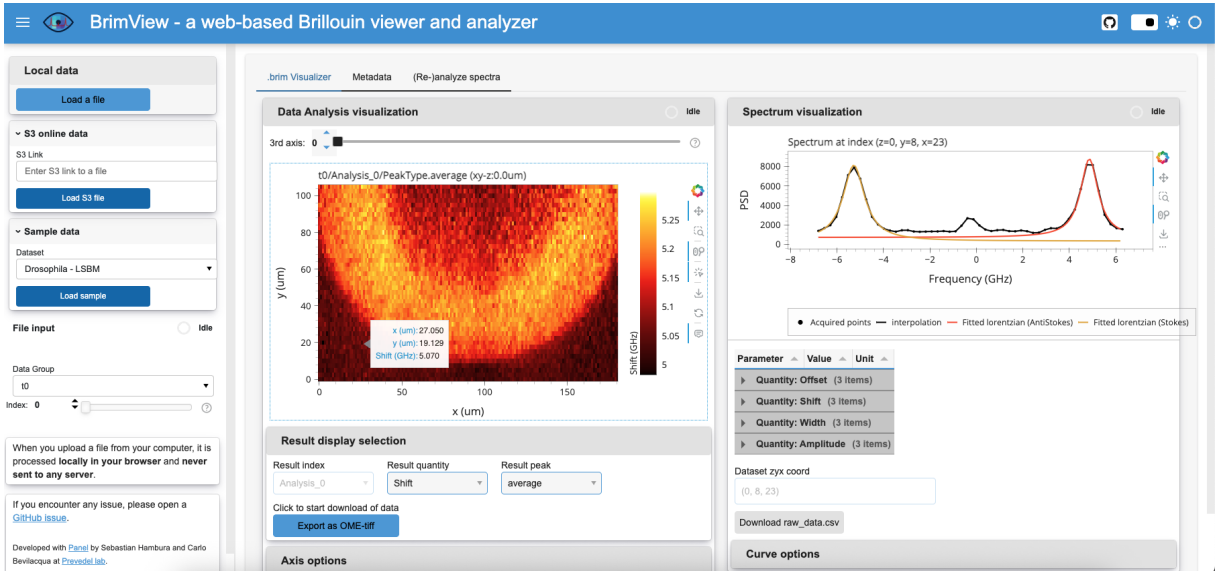

**SI Fig. 2: *BrimView* overview.** The sidebar allows users to load data from local files or S3-compatible URLs. The main interface includes three tabs: (1) Visualization, for browsing image slices and selecting visualization parameters; (2) Metadata, for inspecting acquisition details; and (3) (Re-)analyze spectra to fit all the spectra with different lineshapes/parameters.

As shown in **Supplementary Figure 2**, the BrimView interface consists of a sidebar for file selection (local or S3-based), and three main tabs:

- **Visualization** – Displays 3D Brillouin data as 2D slices. Users can select which quantity (shift, width, …) they want to display. The visualization can be updated by axis orientation, slice index, colormap, and color range. Clicking on any image pixel opens the corresponding spectrum and overlays the fitted curve.
- **Metadata** – Presents structured metadata embedded in the .brim file.
- **PSD Fitting** – This is a custom wrapper around the *Treat* module. It allows us to visualise the mean spectra, and to set the parameters containing the fit and the algorithm. It supports multiple peaks, and various peak models.

|  | Local installation required? | Internet access required? | Performance? | Customizability? |
|---|---|---|---|---|
| **Web version** | ✗ | ✓ *(to load the page)* | ⭐ *(impossible to edit files)* | ⭐ |
| **Standalone application** | ✗ | ✗ | ⭐⭐ | ⭐ |
| **Local Panel server** | ✓ | ✗ | ⭐⭐⭐ | ⭐⭐⭐ |

**SI Table 1: *BrimView* version overview.** Thanks to Panel, the same source code can be used to provide multiple versions of BrimView, tailored to different use-cases.

*BrimView* is implemented using the Python-based **Panel** library, part of the HoloViz ecosystem (ref https://panel.holoviz.org/). It is available in three versions, each tailored to different use cases:

- **Web Version** – Hosted at https://biobrillouin.org/brimview, this version runs entirely in the browser using Pyodide and WebAssembly (WASM). It is ideal for quickly inspecting .brim datasets, e.g., assessing data quality, on any computer without requiring installation.
- **Standalone Application** – A fully self-contained desktop version built using PyInstaller and PyWebview, suitable for offline use on lab workstations without requiring a Python environment. Pre-built binaries and source code are available at https://github.com/prevedel-lab/BrimView/releases.
- **Local Panel Server** – This version runs as a local Python server, offering the highest performance and full access to the underlying code for advanced users. It supports custom workflows, batch processing, and direct modification of the visualization or analysis pipeline. Installation instructions and source code are available at https://github.com/prevedel-lab/BrimView.

*BrimView* is open-source and released under the **LGPL-3.0** license. Contributions, issue reports, and feature requests are welcome via the GitHub repository (https://github.com/prevedel-lab/BrimView/issues).

## Example .brim files

To showcase the capability of the brim file to accommodate data from diverse Brillouin modalities and of the BrimView software to load remote data, we uploaded 4 example dataset on a S3 bucket, under the url https://storage.googleapis.com/brim-example-files:

- drosophila_LSBM.brim.zarr, a 3D dataset of a gastrulating drosophila embryo acquired with a line-scanning spontaneous Brillouin setup[1].
- zebrafish_eye_confocal.brim.zarr, a 3D dataset of a zebrafish eye acquired with a confocal spontaneous Brillouin setup and previously published in Ref. [2].
- zebrafish_ECM_SBS.brim.zarr, a 2D dataset of a zebrafish notochord acquired with a pulsed stimulated Brillouin microscope and published as Figure 3 in Ref. [3].
- oil_beads_FTBM.brim.zarr, a 3D dataset of oil beads embedded in agar, acquired with a FTBM Brillouin microscope and published as Figure 2a and Supplementary Video 1 in Ref. [4].

## BrimX file format for general Brillouin spectroscopy data and local usage

Brillouin scattering serves a variety of applications beyond spatial imaging, some of which may lead to data protection issues. In an effort to unify the storage of data falling in these scenarios, we propose to rely on a HDF5-based format: .brimX.
HDF5 is a robust and well-established file format, widely utilized since 2002 and recognized as a standard across numerous fields of scientific computing [5–7]. The format further comes with a broad spectrum of software platforms (e.g., FIJI, Panoply), can be accessed through existing online tools such as myHDF5, and is supported by various programming languages such as Matlab, Python and R.

This file format being more universal, it is designed to be both human-readable and user-friendly. To facilitate integration and usage within Python environments, a dedicated Python module, *HDF5_BLS*, has been developed to streamline file interfacing and promote efficient creation and utilization of the file format.

### Structure of the BrimX format

The .brimX file format follows a hierarchical logic for storing data. It can succinctly be described as a combination of nested groups, where the last group contains the data in the form of a dataset. The only restriction to the structure of the format we encourage users to have, is that the first group of the file should be named "Brillouin". This will clearly indicate that the elements of this group are specific to BLS measures. This also leaves a potential adjustment of the format to other hyperspectral modalities.

BrimX does not induce restriction neither on the format of its elements nor on their names. To differentiate the nature of the different groups and datasets of the file format, a .brimX-dedicated attribute ("Brillouin_type") is used on every element of the file (whether group or dataset). To get a full description of the file format, please refer to the Preamble section of the Tutorial for the HDF5_BLS project at https://github.com/bio-brillouin/HDF5_BLS/blob/main/guides/Tutorial/Tutorial.pdf.

### Metadata

All metadata are stored as attributes of groups. Attributes are stored hierarchically in the file, meaning that all attributes of parent groups apply by default to all their children. **Supplementary Figure 3** illustrates this logic: a parent group "Brillouin" has a set of attributes that apply to all its children groups. The children groups of the "Brillouin" group have their own set of attributes that only apply to their own elements (in this case a combination of a frequency array and a Power Spectral Density array).

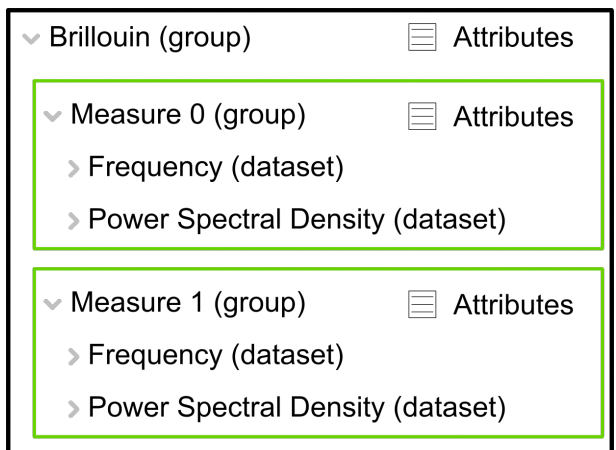

**SI Figure 3**: Illustration of the workings of attributes in the HDF5 Brillouin-specific file format.

The names and values of the metadata are stored as strings in the file format. In an effort to unify the name of the attributes, a spreadsheet is accessible at the repository of the project at https://github.com/bio-brillouin/HDF5_BLS/blob/main/spreadsheets/attributes_v1.0.xlsx. A complete description of the attributes can be found in the Preamble section of the Tutorial for the HDF5_BLS project.

# Integration of the HDF5_BLS library to read/write BrimX files

The .brimX file format has a dedicated library (HDF5_BLS) to interface the file in Python. This library is open-source with a GNU-GPL 3 public license. This library is optimized in memory to handle heavy files. The github repository associated to the library can be found at: https://github.com/bio-brillouin/HDF5_BLS. We especially recommend interested readers to refer to the pdf Tutorial of the library that can be accessed at: https://github.com/bio-brillouin/HDF5_BLS/blob/main/guides/Tutorial/Tutorial.pdf.

## Importing the library and creating a BrimX file

After having installed the HDF5_BLS library from Python Index Package with:
pip install HDF5_BLS
The user can easily create a .brimX file at any location with these lines:
from HDF5_BLS import Wrapper
wrp = Wrapper("path/to/file.h5")# The path where to store the file
Note that this same line of code opens a file located at a given filepath.

## Adding data

Adding a dataset to the file is done by specifying the array to be stored, the location in the file where to store it and the name to give the dataset with functions specific to the type of dataset to be added (e.g. "add_frequency" if a dataset containing the frequency values attributed to a channel are to be stored).

For example for storing a measure coming from an arbitrary spectrometer, the user can use the following line:
wrp.add_raw_data(data, parent_group = "Brillouin/Water Measures/Measure 01", name = "Water spectrum")
Note that the library also comes with a module - HDF5_BLS.load_data - dedicated to importing data and attributes directly from spectrometer files. More information on this module and how to use it can be found in the documentation of the project

## Accessing data

We are relying on a HDF5 file format that can therefore be opened by any HDF5-compatible tool, either online (see https://myhdf5.hdfgroup.org/ for example) or with desktop applications (HDF5view, Panoply or Fiji for example).
Using HDF5_BLS, extracting a dataset located at a given path in the file, and using it in Python, can be done with the following line of code::
data = wrp["Brillouin/Water Measures/Measure 01/Water spectrum"]
In Matlab, this line becomes: data = h5read("path/to/file.h5", "/Brillouin/Water Measures/Measure 01/Water spectrum");

## Integration of library to existing Python workflows

We recommend readers interested in using the proposed approach to just integrate it to their existing workflows. For storing raw data, PSD and frequency axis together with the results of a data processing algorithm for example, the user can use the following code:
############################################
# existing code to import libraries

```
#################################################
from HDF5_BLS import Wrapper
wrp = Wrapper("the/path/where/to/store/the/HDF5/file.h5")

#################################################
# existing code for opening measures
#################################################
wrp.add_raw_data(data, parent_group = "Brillouin/A measure/Acq 1", name = "Raw spectrum") # considering the opened measure is stored at "data" and want to be stored at "Brillouin/A measure/Acq 1" in the file and given the name "Raw spectrum"

#################################################
# existing code for converting to a PSD and frequency array
#################################################
wrp.add_frequency(frequency, parent_group = "Brillouin/A measure/Acq 1", name = "Frequency") # the frequency array is here stored in the "frequency" attribute
wrp.add_PSD(PSD, parent_group = "Brillouin/A measure/Acq 1", name = "Power Spectral Density") # the PSD array is here stored in the "frequency" attribute

#################################################
#existing code for processing the data
#################################################
wrp.add_treated_data(shift = shift, linewidth = linewidth, parent_group = "Brillouin/A measure/Acq 1", name_group = "Treatment 1") # we add here a shift and linewidth array to the file at the same path as the measures, under the "Treatment 1" group
```

Management of metadata

A standard spreadsheet based on the Ref.[8] can be found at https://github.com/bio-brillouin/HDF5_BLS/blob/main/spreadsheets/attributes.xlsx. This spreadsheet can be filled by the user with the values of his own instrument and measures and then applied to any element of the HDF5 file with the following line of code:
wrp.import_properties_data(filepath = "path/to/spreadsheet.xlsx", path = "Brillouin/A measure/Acq 1") # we import here the properties defined in a spreadsheet.

Likewise, it is possible to export a spreadsheet of metadata from a group of HDF5 with the following line of code:
wrp.save_properties_csv(filepath = "path/to/spreadsheet.csv", path = "Brillouin/A measure/Acq 1") # we export all the properties that apply to a specific group or dataset into a csv file.

Note that you can import the properties from an exported file, and that you can also change individual properties with functions of the HDF5_BLS library. Please refer to documentation for details.

Additional features of the BrimX file format
- The library has built-in features to translate raw data obtained from a spectrometer to a physically meaningful Power Spectral Density and Frequency array (see dedicated SI section).
- The package has a built-in feature to perform fitting of the data (see dedicated SI section) that is compatible with both Brim and HDF5-BLS specific file formats.
- To the best of our knowledge, the HDF5 file format cannot host any piece of code that could be executed when opening it. This mitigates most of the risks of direct attacks using HDF5 files and encourages its use. HDF5 files are notorious for not releasing memory after one of their elements has been deleted. We have therefore included tools to re-pack the file on demand.
- .brimX files can be exported from sub-groups of other BrimX files and can be created by combining multiple BrimX files to reduce the size of big files.
- .brimX is essentially a HDF5 file with custom attributes to unify the BLS community. The format is therefore compatible with all HDF5 viewers.
- Users can export datasets in user-friendly formats (2D arrays can be exported as images for example).

## BrimX graphical user interface

The HDF5_BLS library comes with a dedicated Graphical User Interface (GUI) called HDF5_BLS_GUI, to interface it. This GUI is accessible on the same GitHub repository: https://github.com/bio-brillouin/HDF5_BLS.

**Supplementary Figure 4** shows the main window of the HDF5_BLS GUI. The user can either open a BrimX file or drop it to the left pane, where the architecture of the file will be displayed. Each element of the file has an associated icon illustrating its "Brillouin_type" attribute. On the right pane, different tabs allow the user to explore and edit the attributes of any selected element from the right pane and also visualize datasets of 2 dimensions as images. The GUI presents a series of icons on a top container that allow the user to rapidly open an .brimX file, create a new file, save it at a desired location, add data to it, remove data from it, copy a piece of code to extract a selected dataset from the file and export the properties of a selected element to a CSV spreadsheet. By left clicking on any of the elements from the left pane, a context menu appears that allows the user to select between a range of actions to perform on the file element. Finally, all the actions performed in the GUI are logged and displayed on a log text widget located at the bottom of the window.

Here are a non-exhaustive list of what the GUI currently allows:
- The creation of .brimX files by dragging and dropping files, series of files or even directories containing files supported by the HDF5_BLS.load_data module.
- Conversion of data directly obtained from a spectrometer (at the time of the writing: VIPA-based spectrometers, Tandem Fabry-Pérot based spectrometers and Time-resolved based spectrometers) into a standard couple of Power Spectral Density and Frequency arrays that are stored at their expected location in the file.
- Import attributes from a CSV or Excel spreadsheet and export them to a CSV spreadsheet.
- Export the lines of codes to access a given dataset for Python or Matlab workflows.
- Visualize 2D arrays as images, and export them in any format supported by the Python Pillow library.

- Rename groups or datasets, change the value of an attribute or change the "Brillouin_type" attribute of a group or dataset.

The GUI is meant to be used locally and doesn't have the ability to exchange any type of information through the Internet. It is therefore made to comply with all computer safety guidelines both in the public and private sector.

We invite interested readers to refer to the dedicated GitHub repository, and particularly the guides to discover all the features of the GUI and its uses.

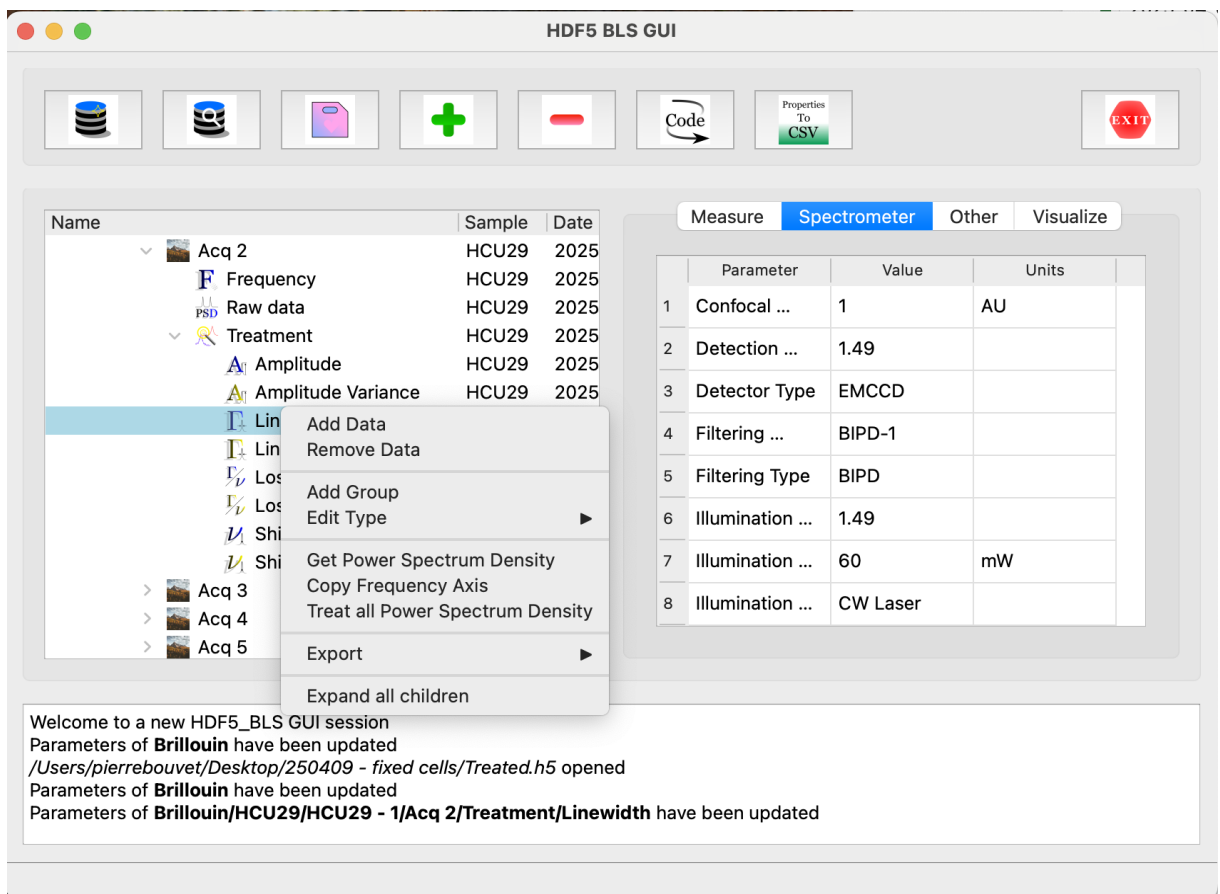

**SI Fig. 4**: An overview of the HDF5_BLS GUI main window where the user has right-clicked on an element.

# Data processing

The project proposes to unify further the data by allowing users to use the same algorithms to obtain a PSD and frequency array from their data. Unifying these transformations presents a double challenge: allowing users to replicate the processing they are used to using with their data while making sure that anyone can replicate their algorithms and examine its steps.

To answer these needs, we have based our approach of the solution on modular algorithms. Rather than defining functions to be called, we have defined classes composed of methods that are used as the individual steps of the algorithm. These classes inherit from low-level classes we have coded to register every call to the methods of a class, and store them in the form of a JSON file. This file stores not only the functions but also the parameters used for these functions, as well as the docstring of these functions.

From there, we propose to differentiate the two main processing steps in extracting information from an arbitrary Brillouin scattering experiment.

## Conversion to a Power Spectral Density

As stated before, we choose to use the Power Spectral Density as the representation of choice of the result of a BLS experiment. The first step in unifying data processing is therefore to convert our results to a physically meaningful doublet of Power Spectral Density and Frequency array. This is done through the HDF5_BLS.analyze module.

This transformation is often spectrometer-specific and we have therefore chosen to define classes specific to the spectrometers that have been used. For instruments that are not yet compatible, we encourage users to refer to the guides available on the HDF5_BLS GitHub repository, a special effort having been put in making the modules developer-friendly.

Additional information as well as tutorials and examples of how to use the HDF5_BLS.analyze module can be found in the documentation of the HDF5_BLS project.

The module is developed to be developer-friendly so we invite anyone interested in participating in the development of this project to refer to the HDF5_BLS GitHub repository and especially the tutorial of the HDF5_BLS package for more information and examples on how to use and expand this module.

## Extraction of information from a Power Spectral Density

To then pass from a PSD and frequency doublet to a value of shift and linewidth, a lineshape fitting has to be performed. As for the conversion of the raw data to a PSD, the lineshape recognition works based on a modular approach. This modular approach allows users to optimize their algorithm and easily compare their efficiency.

The module to process data has further been made independent of the main HDF5_BLS project and can be installed using the Python index package as a single library called "HDF5_BLS_treat" to be used as any other standalone Python library.

This module allows at the moment of writing this article:
- To automatically normalize data (offset recognition and amplitude normalization).
- To fit BLS spectra peak by peak or by fitting a lineshape containing an arbitrary number of peaks.
- To deconvolve peaks by an arbitrary lineshape at the moment of the fit. This deconvolution is performed by fitting the convolution of the peak by the empirically obtained response of the instrument.
- To extract from the fit estimators of errors returned as a variance. This is done by estimating the Hessian matrix from the estimated Jacobian matrix.
- To combine fitted parameters while propagating the error

As for the analysis module, the code is made to be developer-friendly and a dedicated guide on how to implement custom functions can be found on the HDF5_BLS GitHub repository.

## BrimConverter library for .brim ↔ .brimX conversion

*Brimfile* and the *HDF5_BLS* libraries provide the Brillouin microscopy community with organisational, data processing, and visualisation tools that are optimised for both efficiency and flexibility, respectively. Therefore, it is useful to provide a library that acts as an intermediary between the two packages and enables universal compatibility between the two file formats and each of the respective GUI's.

The *BrimConverter* class takes two input filepaths, the 'from' source file containing the Brillouin data, and the 'to' destination file, and a third 'mode' input argument which specifies whether the conversion is between brim→brimX (mode='brim2brimX') or brimX→brim (mode='brimX2brim'). It is worth noting that .brimX files should contain extension .h5 as is consistent with the *HDF5_BLS* convention, and .brim files can contain any of the compatible extension types, e.g., .zarr, .zip, etc.

For the case of brimX→brim conversions (by way of a function called _process_data_group_brimX2brim), e.g., through the below example code.

```python
brimX_file = '/path/to/brimX/file'
brim_file = '/path/to/new/brim/file'
convert_this = BrimConverter(brimX_file, brim_file, mode='brimX2brim')
convert_this.convert()
```

The library invokes an additional class called HDF5Flattener. After loading the input .brimX file, this class parses through the groups looking for "Brillouin_type" attributes, indicating that a group/sub-group is storing the desired Brillouin data (e.g., PSD, frequency, Shift, etc.). One of the merits of the *HDF5_BLS* is the ability to customise one's data storage structure and naming convention. This open-endedness on the front-end, is formalised on the back-end through specification of this "Brillouin_type" attribute and allows the relevant Datasets to be located and scraped for extraction and reformatting into *Brimfile*.

Additionally the utility of *HDF5_BLS* allows Brillouin data to be stored in arrays with arbitrary dimensions/shapes. The HDF5Flattener class accommodates this, and once the relevant Datasets are located with the file structure, they are reshaped into the array-structure enforced by *Brimfile*, e.g., an array containing PSD's should be of shape (z, y, x, f).

The entire .brimX file contents are passed through HDF5Flattener and each unique group containing, e.g., a PSD, its dependent Analysis Datasets (Shift, Linewidth, etc.), and relevant metadata are sequentially passed to a newly created Brimfile through the .create_data_group and .create_analysis_results_group functions resulting in a flattened data structure.

For the case of brim→brimX conversions (through a function called _process_data_group_brim2brimX), e.g., through the below example.

```python
brim_file = '/path/to/brim/file'
brimX_file = '/path/to/new/brimX/file'
convert_this = BrimConverter(brim_file, brimX_file, mode='brim2brimX')
convert_this.convert()
```

The contents of a .brim file are extracted and reformatted into groups and datasets of the designated output brimX.h5 file, including identification of "Brillouin_type" attributes. The *BrimConverter* class can also support two additional input arguments, stop_at=, and map_to=.

The former permits the user to only convert a portion of their dataset (e.g., 3 out of 20 scans), which enables shorter run-times for things like debugging and quickly checking functionality. The map_to= argument accepts either 'cartesian' or 'flat.' The 'flat' option does not reshape the flattened/linearised data from the .brim file, while 'cartesian' remaps/reshapes both PSD and Analysis Datasets back to a z, y, x format for easy visualisation of the output .brimX file in HDF5 viewers or the *HDF5_BLS* GUI.